%% file: main.tex
\documentclass[11pt,a4paper]{article} 
\usepackage[T1]{fontenc} 
\usepackage[utf8]{inputenc} 
\usepackage[english]{babel} 
\usepackage{textcomp} 
\usepackage[english]{varioref} %
\usepackage{float}
\usepackage{comment}
\usepackage{booktabs} 
\usepackage{caption} 
\usepackage{multirow} 
\usepackage{subfig} 
\usepackage{tabularx} 
\usepackage[centertags]{amsmath} 
\usepackage{amsthm} 
\usepackage{amssymb} 
\usepackage{braket} 
\usepackage{tikz} %
\usepackage[numbers, sort&compress]{natbib} 
\usepackage{jheppub} 
\usepackage{todonotes} 
 \usepackage{pgfplots}
\pgfplotsset{compat=1.17}
\usepgfplotslibrary{external}
\usetikzlibrary{pgfplots.groupplots}

\newcommand{\Z}{\mathbb{Z}} 
\newcommand{\varM}{\mathcal{M}} 
\newcommand{\varQ}{\mathcal{Q}} 
\newcommand{\di}{\,\mathrm{d}} 
\newcommand{\zb}{\overline{z}} 
\newcommand{\vev}[1]{\langle\!\langle{#1}\rangle\!\rangle}
\DeclareMathOperator{\gh}{gh} 
\DeclareMathOperator{\MLG}{MLG} 
\renewcommand{\hat}{\widehat} 
\renewcommand{\tilde}{\widetilde} 
\renewcommand{\bar}{\overline} 
\newcommand{\be}{\begin{equation}} 
\newcommand{\ee}{\end{equation}} 

\numberwithin{equation}{section} 
\newsavebox{\riddlebox} 
	{\begin{lrbox}{\riddlebox}
		\begin{minipage}{\dimexpr\columnwidth-2\fboxsep\relax} 
			\textbf{\Large Questions:} \\ \vspace{0em} \itshape}
		{\end{minipage} 
	\end{lrbox}
	\begin{center} 
  \colorbox{lightgray!40!white}{\usebox{\riddlebox}} 
	\end{center} 
	} 
	{\begin{lrbox}{\riddlebox}
		\begin{minipage}{\dimexpr\columnwidth-2\fboxsep\relax} 
			\textbf{\Large To Do:} \\ \vspace{0em} }
		{\end{minipage} 
	\end{lrbox}
	\begin{center} 
  \colorbox{lightgray!40!white}{\usebox{\riddlebox}} 
	\end{center} 
	} 
\title{On construction of correlation numbers in super Minimal Liouville Gravity in the Ramond sector}
\author{V. Belavin, J. Ramos Cabezas, B. Runov}
\affiliation{Physics Department, Ariel University, 
Ariel 40700, Israel.}

\emailAdd{vladimirbe@ariel.ac.il, juanjose.ramoscab@msmail.ariel.ac.il, borisru@ariel.ac.il}

\abstract{We study the construction of correlation numbers in super minimal Liouville gravity. In particular, we construct the fundamental physical fields in the Ramond sector and compute the three-point correlation number involving two physical fields in the Ramond sector and one in the NS sector. Furthermore, we establish the relation between Ramond physical fields and the elements of the ground ring. Using the higher equations of motion of super Liouville theory, this relation leads to a new representation of the Ramond physical fields. This formulation enables a direct analytic computation of correlation numbers involving Ramond field insertions. As an application, we demonstrate the method in the simplest case of a three-point correlation function.}
\begin{document}
\maketitle
\section{Introduction} 

In this paper, we consider $\mathcal{N}=1$ super Minimal Liouville Gravity (SMLG). The bosonic version of MLG has been studied thoroughly, see e.g.~\cite{Zamolodchikov:2005,Belavin:2005yj, Belavin:2006ex,Belavin:2008kv,Belavin:2013nba,Artemev:2022sfi}, the NS sector of SMLG theory has been analyzed in~\cite{Belavin:2008vc,Belavin:2009ag} and references therein. For recent developments, see~\cite{Collier:2023cyw,Collier:2024kwt}. 
In this paper, we focus on the Ramond sector of SMLG.  

There are two alternative approaches to the analysis of two-dimensional gravity: the ``continuous'' approach based on the original definition via the Polyakov action, and the dual ``discrete'' approach based on matrix models.
In the continuous approach, apart from the straightforward calculation of the functional integral, an alternative method allows analytically computing amplitudes using higher equations of motion (HEM) in the underlying Liouville theory. HEMs lead to certain relations for the physical fields which allow analytical calculations of the moduli integrals involved in the definition of the correlation numbers. The corresponding calculations were performed in the Neveu-Schwarz (NS) sector in~\cite{Belavin:2008vc,Belavin:2009ag}.

It was previously conjectured \cite{Seiberg:2003nm} that in the ``discrete'' approach the Ramond sector of the SMLG should be described by the same matrix model as the NS sector. However, this conjecture has not been confirmed as the correlation numbers involving the Ramond fields have never been computed in the ``continuous'' approach. The present study aims to fill this gap at the level of three-point numbers and lays the groundwork for evaluating the moduli integrals appearing in the computation of four-point correlation numbers. We plan to address the latter issue in future work.

Before reporting our results concerning the Ramond sector of SMLG, we find it useful to outline the main issues in the simpler case of the bosonic version of MLG.
MLG is a theory described by three contributions:  a minimal model, $\varM(p,q)$ labelled by two integers $p,q \in \Z$, with central charge $c_{p,q}$; the Liouville CFT, with central charge $c_L$; the theory of ghosts, $(bc)$ system, with central charge $c_{\gh} = -26$.

The action of the theory is
\be
S_{\MLG} = S_{\text{min.mod.}} + S_{\text{Liouville}} + S_{\gh} \; . 
\ee
In MLG one is interested in computing the correlators of physical fields
\be
\vev{\mathbb{V}_1 \mathbb{V}_2 \cdots \mathbb{V}_N}_g \; . 
\label{eqn:1.1}
\ee
Our focus will be on the case of a sphere topology, i.e. a surface of genus $g=0$, and the subscript $g$ will be omitted in what follows. For some results in higher genus topology, see, e.g.,~\cite{Belavin:2010sr,Artemev:2022sfi}.

The fields in~\eqref{eqn:1.1} are constructed as follows 
\be
\mathbb{V}_i(z, \zb) = \phi_{m_i, n_i}(z, \zb) V_{a(m_i, n_i)}(z, \zb)\big(\gh\big)(z, \zb) \; . 
\label{eqn:1.2}
\ee
Here $\phi_{m_i, n_i}(z, \zb)$ is a primary field of the minimal model, parametrized by the integers $m_i$ and $n_i$, $V_{a(m_i, n_i)}$ is the primary field of the Liouville theory, parametrized by the parameter $a$, which depends on the integers $m_i$ and $n_i$, and $\big(\gh\big)$ stands for a composite field made of ghost fields. This form of the fields $\mathbb{V}_i$ of the MLG follows from the fact that they have to represent the BRST cohomology.
In addition to the fields $\mathbb{V}_i$, the set of the physical fields in the model also contains ground ring operators $\mathbb{O}_{m,n}$ \cite{Witten:1991zd, Klebanov:1991hx}.

The definition of the BRST charge $\varQ$ and its nilpotency (or equivalently, the requirement of the Weyl invariance of the theory), furnish also a restriction on the central charges, that has to satisfy 
\be
c_{p,q} + c_L + c_{\gh} = 0 \; \implies \; c_{p,q} + c_L = 26 \; . 
\ee
The correlation function~\eqref{eqn:1.1} is notated with a double angular bracket. This is due to the fact that it is actually defined by an integral over positions of the fields. On the sphere 
\be
\vev{\mathbb{V}_1 \mathbb{V}_2 \cdots \mathbb{V}_N} = \prod_{i=1}^{N-3} \int \di^2 z_i \braket{\prod_{i=1}^{N}\phi_{m_i,n_i}}\braket{\prod_{i=1}^{N}V_{a(m_i,n_i)}} \braket{\gh} \; ,
\label{4point-amplitude-boson-gen}
\ee
where the upper limit $N - 3$ is due to the invariance under global conformal transformation.

In particular, the four-point correlator  is given by
\begin{align}
\Sigma_{MLG}(\{m_i,n_i\})\equiv\vev{\mathbb{V}_1 \mathbb{V}_2 \mathbb{V}_3 \mathbb{V}_4} =  \int \di^2 z \braket{\phi_{m_1,n_1}(z,\bar z)\phi_{m_2,n_2}(0)\phi_{m_3,n_3}(1)\phi_{m_4,n_4}(\infty)}\times\nonumber\\
\times\braket{V_{a(m_1,n_1)}(z,\bar z)V_{a(m_2,n_2)}(0)V_{a(m_3,n_3)}(1)V_{a(m_4,n_4)}(\infty)} \; .
\label{4point-amplitude-boson}
\end{align}
Notice that the contribution of the ghost fields in the case of 4-point amplitude is reduced to the three-point correlation function $\langle c\bar{c}(0) c\bar{c}(1)c\bar{c}(\infty)\rangle=1$.
The crucial step in evaluating the moduli integral in equation (\ref{4point-amplitude-boson}) analytically is to represent (modulo BRST exact terms) the operator $\mathbb{V}_1(z,\bar{z})$ as a Laplacian of the ground ring operator \cite{Belavin:2005yj, Imbimbo:1991ia}
\be
    B_{n_1,m_1}\mathbb{V}_1=\bar{\partial}\partial \bar{H}_{n_1,m_1}H_{n_1,m_1}\phi_{m_1,n_1}V_{n_1,m_1}\bar{c}c+\text{BRST exact terms}\,,
\ee
where the operator $H_{m,n}$ has level $mn$ and ghost number $0$.
This representation allows us to reduce the integral (\ref{4point-amplitude-boson}) to the boundary terms and eventually to compute it completely. Finally, the result is identified with the 4-point number of the matrix model.

This work is dedicated to establishing a connection between physical fields in the Ramond sectors and the correlation numbers in super minimal Liouville gravity (SMLG), with a specific focus on the three-point function. In this case, no integration over the moduli space is required, and the result simplifies to a linear combination of products of three-point functions from each sector.

The general framework of SMLG is reviewed in the next section. Section \ref{secPhys} is devoted to the construction of physical fields in the Ramond sector. In Section 4, we explore the fundamental relation derived from the properties of the ground ring. Our conclusions are presented in Section 5. We note that some of our results appear to be closely related to those in the recent papers ~\cite{Muhlmann:2025ngz,Rangamani:2025wfa}.
\section{Super MLG}
\label{superMLG}
In this section, we review some notations and known results regarding the SMLG. 
The symmetry algebra of SMLG is $\mathcal{N}=1$ superconformal algebra,  
\begin{equation}
\begin{aligned}
\lbrack L_n,L_m]&=(n-m)L_{n+m}+\frac{\hat c}8(n^3-n)\delta_{n,-m},
\\
\{G_r,G_s\}&=2L_{r+s}+\frac{\hat
c}2\left(r^2-\frac14\right)\delta_{r,-s},
\\
[L_n,G_r]&=\left(\frac12n-r\right)G_{n+r},
\end{aligned}
\label{N1algebra}
\end{equation}
where
\begin{equation*}
\begin{alignedat}{2}
&r,s\:\in\mathbb{Z}+\frac12&\quad&\text{for the NS sector},
\\
&r,s\:\in\mathbb{Z}&\quad&\text{for the R sector}.
\end{alignedat}
\end{equation*}

The SMLG~\cite{David:1988hj,Distler:1988jt,Distler:1989nt} is
a tensor product of superconformal matter (SM), super
Liouville \cite{Polyakov:1981re, curtright1984weak, arvis1983spectrum, d1983classical, babelon1984monodromy}, and super ghost systems \cite{Friedan:1985ge, verlinde1989lectures, Polchinski:1998rr, friedan2003tentative}, with the action
\begin{equation}
A_{\text{SLG}}=A_{\text{SM}}+A_{\text{SL}}+A_{\text{SG}}
\end{equation}
each of which obeys the symmetry~\eqref{N1algebra} with  the central
charge parameters constrained  by
\begin{equation}
\hat{c}_{\text{SM}}+\hat{c}_{\text{SL}}+\hat{c}_{\text{SG}}=0\;. \label{totc}
\end{equation}

In the continuous approach, we consider an extension of the minimal Liouville gravity, in which, in the matter sector, we take a generalized minimal model with non-rational values of the central charge $\hat{c}_{SM}$ and non-degenerate primary fields included in the spectrum. The standard constraints on the central charge and fields' conformal dimensions must be restored to compare the results with the matrix model results.
\subsection{Liouville sector}
The central charge is parametrized by  a Liouville coupling parameter $b$
\be
\hat{c}_L=1+2Q^2\,, \quad  Q=b^{-1}+b.
\ee
One can write the primary fields of Super Liouville field theory as exponential vertex operators. For the NS bosonic field $V_a$ and Ramond field $R_a^{\epsilon}$ we have 
\begin{equation} \label{SLprimaries}
    V_a(z)= e^{a \varphi (z)},  \quad R_{a}^{\epsilon}(z)= \Sigma^{\epsilon}  e^{a \varphi(z)},
\end{equation}
where $\Sigma^{\epsilon}$ is the twist field. The conformal dimensions of the primary fields $V_a$ and $R^{\pm}_a$ form a continuous spectrum:
\be
\Delta^{L,NS}(a)=\frac{a(Q-a)}{2}\,,\quad \Delta^{L,R}(a)=\frac{1}{16}+\frac{a(Q-a)}{2}.
\ee
Unless the dimension is equal to $\frac{\hat{c}}{16}$, the highest weight states in super Virasoro modules corresponding to $R_a$ are doubly degenerate.
The pairs of states at the level 0 are interchanged by the action of the fermionic generator $G_0$:
\be
\label{betaL}
    G^{L}_{0}|R^{\sigma}_a\rangle=i\beta^{L}_{a}e^{-\frac{i\pi\sigma}{4}}|R^{-\sigma}_a\rangle\,,\quad
    \beta^{L}_a=\frac{1}{\sqrt{2}}\left(\frac{Q}{2}-a\right)\,.
\ee
The OPE of two nondegenerate NS fields $V_a$ is given by
\be
\begin{split}
    V_{a_1}(x)V_{a_2}(0)=
    \int \frac{dP}{4\pi}|x|^{\Delta-\Delta(a_1)-\Delta(a_2)}
    \Big(&
    \mathbf{C}^{L}_{(a_1)(a_2)(\frac{Q}{2}+iP)}[V_{Q/2+iP}(0)]_{ee}\\
    &+\tilde{\mathbf{C}}^{L}_{(a_1)(a_2)(\frac{Q}{2}+iP)}[V_{Q/2+iP}(0)]_{oo}
    \Big)\,.
\end{split}
\ee
The basic structure
constants $\mathbf{C}^{L}$ and
$\tilde{\mathbf{C}}^{L}$ relevant for NS sector (first computed in ~\cite{Poghossian:1996agj,Rashkov:1996np})
have the explicit form (with $a=a_{1}+a_{2}+a_{3}$)
\begin{equation}
\begin{aligned}
\mathbf{C}_{(a_{1})(a_{2})(Q-a_3)}^{L}&=\kappa_b(a,\mu)\!
\frac{\Upsilon_{\text{R}}(b)\Upsilon_{\text{NS}}(2a_{1})
\Upsilon_{\text{NS}}(2a_{2})\Upsilon_{\text{NS}}(2a_{3})}
{2\Upsilon_{\text{NS}}(a-Q)\Upsilon_{\text{NS}}%
(a_{1+2-3})\Upsilon_{\text{NS}}(a_{2+3-1})\Upsilon_{\text{NS}}(a_{3+1-2})},
\\
\tilde{\mathbf{C}}^{L}_{(a_{1})(a_{2})(Q-a_3)}&=-\kappa_b(a,\mu)
\frac{i\Upsilon_{\text{R}}(b)\Upsilon_{\text{NS}}(2a_{1})
\Upsilon_{\text{NS}}(2a_{2})\Upsilon_{\text{NS}}(2a_{3})}
{\Upsilon_{\text{R}}(a-Q)\Upsilon_{\text{R}}%
(a_{1+2-3})\Upsilon_{\text{R}}(a_{2+3-1})\Upsilon_{\text{R}}(a_{3+1-2})},\\
\kappa_b(a,\mu)&=\left(\!\pi\mu\gamma\!
\left(\frac{Qb}2\right)b^{1-b^{2}}\right)^{\!\!(Q-a)/b}
\end{aligned}
\label{C3}
\end{equation}
where the functions $\Upsilon_{ \mathrm{NS}}(x)$, $\Upsilon_{ \mathrm{R} }(x)$ are defined as follows:
\begin{equation}
\begin{aligned}
\Upsilon_{\text{NS}}(x)&=\Upsilon_{b} \left(\frac
{x}{2}\right)\Upsilon_{b}\left(\frac{x+Q}2\right),
\\
\Upsilon_{\text{R}}(x)&=\Upsilon_{b}
\left(\frac{x+b}2\right)\Upsilon_{b}\left(\frac{x+b^{-1}}2\right)\,.
\end{aligned}
\label{YNSR}
\end{equation}
and the ``Upsilon'' function is defined as  \cite{Zamolodchikov:1995aa}
\begin{equation}
\log \Upsilon_b(x)=\int_0^{\infty} \frac{d t}{t}\left[\left(\frac{Q}{2}-x\right)^2 e^{-t}-\frac{\sinh ^2\left(\frac{Q}{2}-x\right) \frac{t}{2}}{\sinh \frac{b t}{2} \sinh \frac{t}{2 b}}\right].
\end{equation}

The 3-point functions in N=1 SLFT involving two fields from the Ramond sector 
and  the superfield from the NS sector have the following structure \cite{Poghossian:1996agj}
\be \label{apOPERRV}
\begin{split}
    \Big< R^{\epsilon_1}_{\alpha_1}(z_1)R^{\epsilon_2}_{\alpha_2}(z_2)
    \Big(V_{\alpha_3}(z_3)+&\theta G_{-\frac{1}{2}}V_{\alpha_3}(z_3)+\bar{\theta}\,\bar{G}_{-\frac{1}{2}}V_{\alpha_3}(z_3)+\theta\bar{\theta}G_{-\frac{1}{2}}\bar{G}_{-\frac{1}{2}}V_{\alpha_3}(z_3)\Big)\Big>
    \\
    =
  v_{3pt}(z_1,z_2,z_3)
       &
       \left(
    \delta_{\epsilon_1,\epsilon_2} 
    \left(\mathbf{C}^{L,\epsilon_1}_{[\alpha_1][\alpha_2](\alpha_3)}+\frac{|z_{12}|\theta\bar{\theta}}{|z_{13}z_{23}|}\tilde{\mathbf{C}}^{L,\epsilon_1}_{[\alpha_1][\alpha_2](\alpha_3)}\right)\right.
    \\
    &
    \left.
    +\delta_{\epsilon_1,-\epsilon_2}\left(\frac{z_{12}^{\frac{1}{2}}\theta}{(z_{13}z_{23})^{\frac{1}{2}}} \mathbf{d}^{L,\epsilon_1}_{[\alpha_1][\alpha_2](\alpha_3)}+\frac{\bar{z}_{12}^{\frac{1}{2}}\bar{\theta}}{(\bar{z}_{13}\bar{z}_{23})^{\frac{1}{2}}}\,\bar{\mathbf{d}}^{L,\epsilon_1}_{[\alpha_1][\alpha_2](\alpha_3)}\right)
    \right)\\
\end{split}
\ee
with
\be
  v_{3pt}(z_1,z_2,z_3)=|z_{12}|^{-2\gamma_{123}}
     |z_{13}|^{-2\gamma_{132}}
      |z_{23}|^{-2\gamma_{231}}
\ee
\be
    \gamma_{ijk}=\Delta(\alpha_i)+\Delta(\alpha_j)-\Delta(\alpha_k)\,,\quad z_{ij}=z_i-z_j\,.
\ee
All the structure constants appearing in (\ref{apOPERRV}) ($\mathbf{C}, \tilde{\mathbf{C}}, \mathbf{d}, \bar{\mathbf{d}}$) can be expressed in terms of two structure constants $\mathbf{C}^{L,\pm}_{[\alpha_1][\alpha_2](\alpha_3)}$,
which are given by \cite{Poghossian:1996agj}
\begin{equation} \label{Ramondstructconst}
\begin{aligned}
\mathbf{C}_{[a_1][a_2] (a_3)}^{L,\epsilon} & =\frac{1}{2}\left(\pi \mu\gamma\left(\frac{Q b}{2}\right) b^{1- b^2}\right)^{\frac{Q-a}{b}} \\
& \times\left[\frac{\Upsilon_{\mathrm{R}}(b) \Upsilon_{\mathrm{R} }\left(2 a_1\right) \Upsilon_{\mathrm{R}}\left(2 a_2\right) \Upsilon_{\mathrm{NS}}\left(2 a_3\right)}{\Upsilon_{\mathrm{R}}(a-Q) \Upsilon_{\mathrm{R}}\left(a_1+a_2-a_3\right) \Upsilon_{\mathrm{NS}}\left(a_2+a_3-a_1\right) \Upsilon_{\mathrm{NS}}\left(a_3+a_1-a_2\right)}\right. \\
& \left.+\epsilon \frac{\Upsilon_{  \mathrm{R} }(b) \Upsilon_{\mathrm{R}}\left(2 a_1\right) \Upsilon_{\mathrm{R}}\left(2 a_2\right) \Upsilon_{\mathrm{NS}}\left(2 a_3\right)}{\Upsilon_{\mathrm{NS}}(a-Q) \Upsilon_{\mathrm{NS}}\left(a_1+a_2-a_3\right) \Upsilon_{\mathrm{R}}\left(a_2+a_3-a_1\right) \Upsilon_{\mathrm{R}}\left(a_3+a_1-a_2\right)}\right].
\end{aligned}
\end{equation}
This normalization is such that the reflection coefficients (the two-point functions) are given by
\begin{equation}
\frac{\mathbf{C}_{[a][a_2] (a_3)}^{L,\epsilon}}{\mathbf{C}_{[Q-a][a_2] (a_3)}^{L,\epsilon}}=G_R(a)=\frac{\gamma \left(a b-\frac{b Q}{2}+\frac{1}{2}\right) \left(\pi  u \gamma \left(\frac{1}{2} \left(b^2+1\right)\right)\right)^{\frac{Q-2 a}{b}}}{\gamma \left(-\frac{a}{b}+\frac{Q}{2 b}+\frac{1}{2}\right)},
\end{equation}
\begin{equation} \label{refcoefRN}
\frac{\mathbf{C}_{L,[a_1][a_2] (a)}^\epsilon}{\mathbf{C}_{L,[a_1][a_2] (Q-a)}^\epsilon}=  G_{NS}(a)=\frac{b^2 \gamma \left(a b-\frac{b Q}{2}\right) \left(\pi  u \gamma \left(\frac{1}{2} \left(b^2+1\right)\right)\right)^{\frac{Q-2 a}{b}}}{\gamma \left(\frac{Q}{2 b}-\frac{a}{b}\right)}\,,
\end{equation}
where we use the standard notation
\begin{equation} \label{littlegamma}
    \gamma(x)= \frac{\Gamma(x)}{\Gamma(1-x)},
\end{equation}
and $\Gamma(x)$ is the gamma function. A special role is played by a discrete set of the degenerate fields $V_{m,n}$ and $R^{\pm}_{m,n}$ satisfying
\be
\label{SLdeg}
    V_{m,n}=V_a\Big|_{a=a_{m,n}}\,,\quad
    R^{\pm}_{m,n}=R^{\pm}_a\Big|_{a=a_{m,n}}\,,
    \quad a_{m,n}=\frac{Q}{2}-\frac{b^{-1}m+bn}{2},
\ee
\be
    D^{L}_{m,n}V_{m,n}=0\; (m-n \mod 2 =0)\,,\quad
    D^{L}_{m,n}R^{\pm}_{m,n}=0\; (m-n \mod 2 =1)\,.
\ee
Below we will require also the special structure constants
\begin{equation} \label{specialstrc}
\begin{aligned}
& \mathbf{C}_{[-b / 2] [a+b / 2](a)}^{\epsilon}=\epsilon\left[\frac{\gamma(Q b) \gamma\left(\frac{(2 a-Q) b}{2}\right)}{\gamma\left(\frac{Q b}{2}\right) \gamma(a b)}\right]^{1 / 2}, \\
& \mathbf{C}_{[-b / 2] [a-b / 2](a)}^{\epsilon}=\left[\frac{\gamma(Q b) \gamma\left(\frac{(Q-2 a) b}{2}\right)}{\gamma\left(\frac{Q b}{2}\right) \gamma((Q-a) b)}\right]^{1 / 2} .
\end{aligned}
\end{equation}
Notice that (\ref{specialstrc}) uses a different normalization (which corresponds to the normalization that we use for the matter sector), since $ \mathbf{C}_{[-b / 2] [-b / 2](0)}^{\epsilon}=1$ (we omit the index ``L'' to emphasize this).

Following \cite{Zamolodchikov:2003yb,Belavin:2006pv} we introduce logarithmic degenerate fields
\be
    V_{m,n}^{\prime}(x)=\frac{\partial}{\partial a}V_{a}(x)\Big|_{a=a_{m,n}}\,,\quad
    R_{m,n}^{\pm\prime}(x)=\frac{\partial}{\partial a}R^{\pm}_{a}(x)\Big|_{a=a_{m,n}}\,,
\ee
which are subject to Higher Equations of Motion
\be
\label{HEM}
\bar{D}^{L}_{m,n}D^{L}_{m,n}V^{\prime}_{m,n}=B_{m,n}V_{m,-n}\,,\quad
    \bar{D}^{L}_{m,n}D^{L}_{m,n}R^{\pm\prime}_{m,n}=B_{m,n}R^{\pm}_{m,-n}.
\ee
The factors  $B_{m,n}$ in the NS sector are given by
\begin{equation} \label{bmnfactor}
    B_{m, n}=2^{m n} i^{m n-2[m n / 2]} b^{n-m+1}[\pi \mu \gamma(b Q / 2)]^n \gamma\left(\frac{m-n b^2}{2}\right) \prod_{(k, l) \in\langle m, n\rangle_{\mathrm{NS}}} \lambda_{k, l},
\end{equation}
and in the Ramond sector
\begin{equation}
    B_{m, n}=2^{m n} b^{n-m}[\pi \mu \gamma(b Q / 2)]^n \gamma\left(\frac{1}{2}+\frac{m-n b^2}{2}\right) \prod_{(k, l) \in\langle m, n\rangle_{\mathrm{R}}} \lambda_{k, l},
\end{equation}
where
\begin{equation}
    \lambda_{m,n}=  \frac{m b^{-1}+ n b}{2},
\end{equation}
and the sets $\langle m, n\rangle_{\mathrm{NS}}$, $\langle m, n\rangle_{\mathrm{R}}$ are defined as
\begin{equation}
\begin{aligned}
& \langle m, n\rangle_{\mathrm{NS}}=\{1-m: 2: m-1,1-n: 2: n-1\} \\
& \cup\{2-m: 2: m-2,2-n: 2: n-2\} \backslash\{0,0\}, \\
& \langle m, n\rangle_{\mathrm{R}}=\{1-m: 2: m-1,2-n: 2: n-2\} \\
& \cup\{2-m: 2: m-2,1-n: 2: n-1\} \backslash\{0,0\}
\end{aligned}
\end{equation}
where as usual $a:d:b$ stands for the sets {$a,a+d,a+2d,...,b$ (from $a$ to $b$ with step $d$).

\subsection{Matter sector}
The role of conformal matter in SMLG is played by a supersymmetric generalized minimal model (GMM). Contrary to the standard definition of a minimal model, the central charge can assume nonrational values.
The primary fields of this model in the NS sector are $\Phi_{\alpha}$ (where $\alpha$ is a continuous parameter), and the Ramond sector comprises the fields $\Theta_{\alpha}^{\pm}$ .
GMM can be obtained from the supersymmetric Liouville theory by  performing the transformations 
\be
\label{subsML}
    b\to  ib\,,\quad a \to -i a\,,
\ee
though we must choose a different normalization
\be
    G_{NS}^{M}(\alpha)=G_{R}^{\alpha}=1\,.
\ee
The primary fields corresponding to degenerate representations are labeled by pairs of integers $m,n$ (where $m-n$ is even for NS fields and odd for Ramond fields). They form a closed fusion algebra and satisfy
\be
    D^{M}_{m,n}\Phi_{m,n}=0\,,\quad
    D^{M}_{m,n}\Theta_{m,n}^{\pm}=0\,.
\ee
To fulfill~\eqref{totc}, the parameter $b$ must be related to the central charge of the matter theory as follows
\be
\hat{c}_M=1-2q^2\,, \quad  q=b^{-1}-b.
\ee
Thus, the conformal dimensions of the primary fields in the NS and Ramond sectors can be written respectively as
\be
\Delta^{M,NS}(\alpha)=\frac{\alpha(\alpha-q)}{2}\,,\quad \Delta^{M,R}(\alpha)=\frac{1}{16}+\frac{\alpha(\alpha-q)}{2},,
\ee
and the conformal dimensions of degenerate fields read
\be
    \Phi_{n,m}=\Phi_{\alpha_{n,m}}\,,\quad
    \Theta_{n,m}^{\pm}=\Theta_{\alpha_{n,m}}^{\pm}\,,
    \quad \alpha_{n,m}=\frac{q}{2}-\lambda_{-m,n}\,.
\ee
The fields $\Theta^{-}_{\alpha}$ are normalized so that the parameters $\beta^{M}_\alpha$ defining the action of the generator $G_0^{M}$ are related to analogous parameters $\beta^{L}$ in the Liouville sector as follows:
\be
\label{betaM}
    G_0^{M}|\Theta_{\alpha}^{\sigma}\rangle=
    i\beta^{M}_{\alpha}e^{-\frac{i\pi\sigma}{4}}
    |\Theta_{\alpha}^{-\sigma}\rangle\,,\quad
    \beta^{M}_{a-b}=-i\beta^{L}_a\,.
\ee
The matter special structure constants are easily obtained from (\ref{specialstrc}) by performing the substitution (\ref{subsML}), obtaining 
\begin{equation}
\begin{aligned}
& \mathbf{C}^{M,\epsilon}_ {  [b/2] [a] (a-b/2)   }   = \left(\frac{\gamma \left(1-b^2\right) \gamma \left(\frac{1}{2}-a b\right)}{\gamma \left(\frac{1}{2}-\frac{b^2}{2}\right) \gamma \left(-a b-\frac{b^2}{2}+1\right)}\right)^{1/2}, \\ &
\mathbf{C}^{M, \epsilon}_ {  [b/2] [a] (a+b/2)   }  = \epsilon \left(\frac{\gamma \left(1-b^2\right) \gamma \left(a b+b^2-\frac{1}{2}\right)}{\gamma \left(\frac{1}{2}-\frac{b^2}{2}\right) \gamma \left(\frac{1}{2} b (2 a+b)\right)}\right)^{1/2}.
\end{aligned}
\end{equation}
\subsection{Ghost sector}
The super ghost system is described by the free superconformal field theory with the central charge
\be
\hat{c}_{gh}=-10.
\ee
It contains  two anticommuting fermionic fields $(b,c)$ with spins $(2,-1)$ and the bosonic fields $(\beta,\gamma)$ with spins
$(3/2,-1/2)$. The fields (see~\cite{Belavin:2008vc}) of the form
$\delta(\gamma(0))$ of dimension $1/2$ arise in the construction of the amplitudes in NS sector. The basic properties of these fields are
\begin{equation}
    b(z)= \sum_n \frac{b_n}{z^{n+2}}, \quad   c(z)= \sum_n \frac{c_n}{z^{n-1}}  ,\quad   \beta(z)= \sum_n \frac{\beta_n}{z^{n+3/2}} ,\quad \gamma = \sum_{n}  \frac{\gamma_n}{z^{n+1/2}},
\end{equation}
\begin{equation} \label{bcgbetaope}
    b(z)c(0) =\frac{1}{z}, \quad \gamma(z) \beta(0) = \frac{1}{z}.
\end{equation}
The modes of the fermionic ghosts $b_n,c_n$ are labeled by integers.
The modes of the bosonic ghosts $\beta_k,\gamma_k$ are labeled by half-integers in the NS sector and by integers in the R sector. The stress-energy tensor $T^{g}$ and its superpartner $G^{g}$ can be written as follows (normal ordering is implied in the definitions)
\begin{equation}
\begin{aligned}
  &  T^g=     -2 b c' - b'c - \frac{3}{2} \beta \gamma' -  \frac{1}{2}\beta' \gamma ,\\ &
    G^{g} =   \beta' c +\frac{3}{2}\beta c' -2 b \gamma  .
\end{aligned}
\end{equation}
Their super Virasoro generators are given by 
\begin{align}
\label{ghost-Vir-gener}
&L_m^{g}=\sum_n(m+n):b_{m-n} c_n: +\sum_k(\frac{m}2+k):\beta_{m-k} \gamma_k:  + a^{gh} \delta_{m,0}  ,\\
&G_k^{g}=-\sum_n\left[(k+\frac{n}2):\beta_{k-n} c_n:+2 b_n\gamma_{k-n}\right],
\end{align}
where $a^{gh}=-1/2$ in the NS sector and $a^{gh}=-5/8$ in the R sector.
\section{Physical fields}
\label{secPhys}
\subsection{BRST charge}
SMLG is a BRST invariant theory with the (nilpotent) BRST charge $\mathcal{Q}$ given by
\begin{equation} \label{BRSTcharge}
\begin{aligned}
  &  \mathcal{Q} = \frac{1}{2 \pi i} \oint dz j_{\mathcal{Q}} (z)  ,\\&
  j_{\mathcal{Q}}(z)= :c(z) \left( T^{L}(z)+T^{M}(z) +\frac{1}{2} T^{g}(z) \right) +
  \gamma(z) \left( G^{L}(z)+G^{M}(z)+ \frac{1}{2} G^{g}  (z)\right) :,
\end{aligned}
\end{equation}
 and it obeys the nilpotent relation $  \mathcal{Q}^2=0$. The charge can also be written in terms of modes
\begin{align}
\mathcal{Q}=\sum_m{:}\bigg[L_m^{\text{M+L}}+\frac{1}{2}L^{\text{g}}_m\bigg]c_{-m}{:}+
\sum_r{:}\bigg[G_r^{\text{M+L}}+\frac{1}{2}G^{\text{g}}_r\bigg]\gamma_{-r}{:}+\frac{a^{gh}}{2}c_0,
\label{Q}
\end{align}
which obey the following commutation relations with the superconformal generators

\begin{align}
\{b_n,\mathcal{Q}\}=L^{L}_n+L^{M}_n+L^{g}_n,
\label{b-Q-comm}\\
\{\beta_r,\mathcal{Q}\}=G^{L}_r+ G^{M}_r+ G^{g}_r.
\label{beta-Q-comm}
\end{align}
The physical fields form a space of $\mathcal{Q}$-cohomology classes. From~\eqref{b-Q-comm} it follows that the total conformal dimension of any physical field $\Psi$ is zero
\be
\Delta^{tot}(\Psi)=0.
\label{delta-tot}
\ee
\subsection{Physical fields in the NS sector}
In NS sector, there exist two types of physical fields
\begin{align} 
\mathbb{W}_{a}(z,\bar z)=\mathbb{U}_{a}(z,\bar z)\cdot c(z)\bar c(\bar
z)\cdot \delta(\gamma(z))\delta(\bar\gamma(\bar z)), \label{W}
\end{align}
and
\begin{align}
\tilde{\mathbb{W}}_{a}(z,\bar z)=\biggl(\bar G^{\text{M+L}}_{-1/2}+
\frac12\bar G_{-1/2}^{\text{g}}\biggr)\biggl(G^{\text{M+L}}_{-1/2}+
\frac12G_{-1/2}^{\text{g}}\biggr)\mathbb{U}_{a}(z,\bar z)\cdot\bar
c(\bar z)c(z), \label{tildeW}
\end{align}
where
\begin{align}
\mathbb{U}_{a}(z,\bar z)=\Phi_{a-b}(z,\bar z)V_{a}(z,\bar z).
\end{align}
The parameter $a$ can take generic values. The general form of
the $n$-point correlation numbers on the sphere for these
observables~\cite{Belavin:2008vc} is
\begin{equation}
I_n(a_1,\cdots,a_n)=\prod_{i=4}^{n}\int d^2 z_i \biggl\langle \bar
G_{-1/2}G_{-1/2}\mathbb{U}_{a_i}(z_i)
\tilde{\mathbb{W}}_{a_{1}}(z_{1})\,\mathbb{W}_{a_{2}}(z_{2})\,
\mathbb{W}_{a_{3}}(z_{3})\biggr\rangle. \label{corrSMG}
\end{equation}
In particular, the three-point correlation number reads
\be \label{purNSI3}
    I_3(a_1,a_2,a_3)=\langle \tilde{\mathbb{W}}_{a_1}(z_1) 
    \mathbb{W}_{a_2}(z_2)
    \mathbb{W}_{a_3}(z_3)
    \rangle=
    \Omega_{NS}(b)\prod\limits_{i=1}^{3}N_{NS}(a_i),
\ee
with the ``leg'' factors $N_{NS}$ given by
\begin{equation}
N_{NS}(a)=\biggl[\pi\mu\gamma\biggl(\frac12+\frac{b^2}2\biggr)\biggr]^{-a/b}
\biggl[\gamma\biggl(ab-\frac{b^2}2+\frac12\biggr) \gamma\biggl(\frac
ab-\frac{b^{-2}}2+\frac12\biggr)\biggr]^{1/2}, \label{N}
\end{equation}
and the $a$-independent overall factor is 
\begin{equation}
\Omega_{NS}(b)=i\left[\pi \mu \gamma\left(\frac{1}{2}+\frac{b^2}{2}\right)\right]^{Q / b}\left[\frac{\gamma\left(b^2 / 2+1 / 2\right) \gamma\left(b^{-2} / 2-1 / 2\right)}{b^2}\right]^{1 / 2}.
\end{equation}
To facilitate the comparison of our results with ``discrete'' approach we need expressions normalized by partition function of the super Liouville theory.
The latter can be restored from the three-point function (\ref{purNSI3}):
\be
    I_3(b,b,b)=-i\partial^{3}_{\mu}Z_L
\ee
\be
    Z_L=
    \frac{b^{6}\mu^{\frac{Q}{b}}}{b^4-1}
    \left[\pi\gamma\left(\frac{1}{2}+\frac{b^2}{2}\right)\right]^{\frac{Q}{b}-3}
    \gamma^{2}(b^2/2+1/2)\gamma(3/2-b^{-2}/2)
\ee

For the analytic computation of the amplitudes~\eqref{corrSMG} we introduce the ground ring physical states, constructed as follows
\begin{equation}
\mathbb{O}_{m,n}(z,\bar z)=\bar H_{m,n}H_{m,n}\Phi_{m,n}(z,\bar
z)V_{m,n}(z,\bar z).
\end{equation}
The operators $H_{m,n}$ are composed of the super Virasoro
generators and are defined uniquely modulo $\mathcal{Q}$-exact terms. 

For the analytic calculation there are two important relations~\cite{Belavin:2008vc}
\begin{equation}
  \bar{\mathcal{Q}} \mathcal{Q} \mathbb{O}'_{m,n}=B_{m,n}\tilde{\mathbb{W}}_{m,-n} \label{basic},
\end{equation}
and
\begin{align}
\bar G_{-1/2}G_{-1/2}\mathbb{U}_{m,-n}=
B_{m,n}^{-1}\bar\partial\partial \mathbb{O}'_{m,n}\mod \mathcal{Q},
\label{GGU1}
\end{align}
where the logarithmic counterparts of the discrete states
$\mathbb O_{m,n}$,
\begin{equation}
\mathbb{O}'_{m,n}=\bar H_{m,n}H_{m,n}\Phi_{m,n}V_{m,n}',
\end{equation}
and $B_{m,n}$ are the coefficients (\ref{bmnfactor}) arising in the higher equations
of motion of SLFT~\cite{Belavin:2006pv}. 

\subsection{Construction of the physical fields in the R sector}
In this section we describe the construction of local physical fields in the Ramond sector of SMLG.
Some discussion of the Ramond sector of SMLG can be found in~\cite{Distler:1989nt}. Physical fields $|\Psi\rangle$ satisfy the following requirements
\begin{align}
\label{PhysCond}
    &\mathcal{Q}|\Psi\rangle=0,\quad |\Psi\rangle\neq \mathcal{Q} [...],\\
&b_0|\Psi\rangle=L_0|\Psi\rangle=0.
\end{align}
In the Ramond sector there is an additional constraint 
\be
\beta_0|\Psi\rangle=G_0|\Psi\rangle=0.
\label{beta0constr}
\ee
As well as in the NS sector we construct $|\Psi\rangle$ from the primary fields in the Matter and Liouville sector
\be
\label{Uansatz}
|\mathbb{U}^{R}_a\rangle =\sum_{\epsilon,\epsilon^{\prime}=\pm 1}u^{R}_{\epsilon,\epsilon^{\prime}}|\Theta^{\epsilon}_{a-b}\rangle|R^{\epsilon^{\prime}}_{a}\rangle.
\ee
Its dimension is  
\be
\Delta(\mathbb{U}^{R}_a)=5/8.
\label{dimU-ramond}
\ee
In order to fulfill the condition $b_0|\Psi\rangle=0$, in the $(bc)$ sector, the state $|\Psi\rangle$ must contain a vacuum $|v\rangle_{bc}$ according to
\be
\label{bc_vac}
\begin{cases}
    b_n|v\rangle_{bc}=0,\quad n\geq0,\\
    c_m|v\rangle_{bc}=0,\quad m\geq1.
\end{cases}
\ee
The state $|v\rangle_{bc}$ corresponds to the field $c(z)$ with conformal dimension $\Delta(c)=-1$.

As ingredient of $|\Psi\rangle$ in the $(\beta,\gamma)$ sector we have to take a state $|v\rangle_{\beta\gamma}$, consistent with~\eqref{beta0constr} and zero total dimension condition~\eqref{delta-tot}
\be \label{cdzeroeq}
5/8- 1+ \Delta(|v\rangle_{\beta\gamma})=0,
\ee
so that
\begin{align}
&L_0|v\rangle_{\beta\gamma}=\frac{3}{8}|v\rangle_{\beta\gamma},\\
&\beta_0 |v\rangle_{\beta\gamma}=0.
\end{align}
From the explicit form of the generator $L_0$ (see~\eqref{ghost-Vir-gener}) we get
\be
\label{RghVacBoson}
\begin{cases}
    \beta_n|v\rangle_{\beta\gamma}=0,\quad n\geq0,\\
\gamma_m|v\rangle_{\beta\gamma}=0,\quad m\geq1.
\end{cases}
\ee
Notice, that from the general classification it follows that in the picture $q$
\be
\begin{cases}
    \beta_n|v,q\rangle_{\beta\gamma}=0,\quad n\geq-q-1/2,\\
\gamma_m|v,q\rangle_{\beta\gamma}=0,\quad m\geq q+3/2.
\end{cases}
\ee
Thus, the appropriate vacuum state corresponds to the highest state $|v,-1/2\rangle$ in picture $q=-1/2$. We shall denote the corresponding field as $\sigma(z)$.
The physical field in the Ramond sector is therefore given by
\be
\label{RamondPhys}
    \mathbb{R}_a=\mathbb{U}_a^{R}\bar{c}c\bar{\sigma}\sigma.
\ee
To fix the field completely, it remains to determine the coefficients $u_{\epsilon,\epsilon^{\prime}}^{R}$ in the equation (\ref{Uansatz}). Upon substitution of (\ref{bc_vac}) and (\ref{RghVacBoson}) the condition (\ref{beta0constr}) reduces to\footnote{Similarly, for the antiholomorphic part $   \bar{G}^{L+M}_0|\mathbb{U}_a^{R}\rangle=0.$}
\be 
\label{G0condLM}
    G^{L+M}_0|\mathbb{U}_a^{R}\rangle=0.
\ee
On the other hand, in order for the Ramond sector to be described by the same matrix model as NS case, the three-point number
\be \label{IRRN1}
    I_{RRN}(a_1,a_2,a_3)=\langle \mathbb{R}_{a_1}\mathbb{R}_{a_2}\mathbb{W}_{a_3}\rangle
\ee
must be factorizable into a product of ``leg'' factors depending only on respective Liouvlle momenta $a_i$ similar to the correlation number of three NS fields (\ref{purNSI3}).

Using explicit form of the action of generators $G^{L}_0,G^{M}_0$ (\ref{betaL}),(\ref{betaM}) and considerations of Appendix \ref{apShiftRelation}} we find the only solution compatible with factorizability of the three-point number 
\be \label{URoption2}
\mathbb{U}^{R}_a=\Theta_{a-b}^{-}R_{a}^{+}+i\Theta_{a-b}^{+}R_a^{-}.
\ee
The three-point number (\ref{IRRN1}) then evaluates to
\be
\label{IRRN_II}
\begin{split}
    I_{RRN}=&\frac{\mathbf{X}^{-}_{MLG}(a_1,a_2|a_3)-\mathbf{X}^{+}_{MLG}(a_1,a_2|a_3)}{2}\\
    &=\mathbf{C}_{[a_1][a_2](a_3)}^{L,+}\mathbf{C}_{[a_1-b][a_2-b](a_3-b)}^{M,-}-\mathbf{C}_{[a_1][a_2](a_3)}^{L,-}\mathbf{C}_{[a_1-b][a_2-b](a_3-b)}^{M,+}.
\end{split}
\ee
Substituting explicit expressions for the three-point structure constants in Liouville and matter sector into (\ref{IRRN_II}) we obtain the following factorization for the SMLG three-point correlation number
\be \label{3ptnumber}
       \langle \mathbb{R}_{a_1}\mathbb{R}_{a_2}\mathbb{W}_{a_3}\rangle=\Omega_{R}(b)N_R(a_1)N_R(a_2)N_{NS}(a_3),
\ee
where we found that the Ramond leg factor $N_R(a)$ is given by
\begin{equation} \label{NRlegfactor}
    N_R(a)=\sqrt{\gamma \left(\frac{a}{b}-\frac{1}{2 b^2}\right) \gamma \left(a b-\frac{b^2}{2}\right)} \left(\pi  \mu \gamma \left(\frac{1}{2} \left(b^2+1\right)\right)\right)^{-\frac{a}{b}}\,.
\end{equation}
The NS-sector leg factor $N_{NS}(a_3)$ coincides with the expression (\ref{N}), despite the fact that the structure constants in the NS and Ramond sectors, given in (\ref{C3}) and (\ref{Ramondstructconst}) respectively, are not identical. The overall factor $\Omega_R(b)$ is determined by evaluating (\ref{IRRN_II}) for a specific configuration, for instance $(a_1, a_2, a_3) = (3b/2, 3b/2, b)$, and reads
\begin{equation} \label{ROmegafactor}
  \Omega_R(b)=  \frac{ b^{-3} \left( \pi \mu \right) ^{\frac{1}{b^2}+1} \sqrt{\gamma \left(\frac{1}{2}-\frac{1}{2 b^2}\right)} \gamma \left(\frac{1}{2} \left(b^2+1\right)\right)^{\frac{1}{b^2}+2}  }{\gamma \left(\frac{3}{2}-\frac{1}{2 b^2}\right) \sqrt{\gamma \left(\frac{1}{2} \left(b^2-1\right)\right)}}.
\end{equation}
The derivation of the factorized formula (\ref{3ptnumber}) and the explicit expression (\ref{NRlegfactor}) is presented in Appendix~\ref{apShiftRelation}. The NS leg factor $N_{NS}(a_3)$ is obtained by computing the analogous shift relations, namely (\ref{apeshiftrela4}), (\ref{xmratDeltaN}), and (\ref{apRshiftRela}), for the NS field appearing in (\ref{apeshiftrela1}). This can be done either by directly applying the reasoning outlined in the appendix to the NS case, or by using the expressions of the structure constants (\ref{Ramondstructconst}). The results (\ref{3ptnumber}), (\ref{NRlegfactor}), and (\ref{ROmegafactor}), together with the expression for the NS leg factor, represent an important part of the main results of this work.
\section{Three-point numbers involving Ramond fields}
\subsection{Ghost number balance rules} \label{ghostnumberrules}
The conservation of the fermionic ghost current restricts the types of fields that can appear in non-vanishing correlation functions of SMLG. Recall that, in the NS sector, there are two ghost balance rules \cite{Belavin:2008vc}
\begin{equation} \label{gbalance1}
    N_c-N_b=3,
\end{equation}
\begin{equation} \label{gbalance2}
        N_{\delta(\gamma)} - N_{\delta (\beta)}+ N_{\beta}- N_{\gamma}=2.
\end{equation}
Before considering analogous rules required in the Ramond sector, let us provide a brief overview of the derivation of these rules. For $bc$ system, in the correlation function involving only physical fields with $N_c$ and $ N_b$ numbers of $c$ and $b$ fields respectively, we insert the operator $N^{g}_{bc}= \oint dz J^{bc}$, where $J^{bc}= -: bc:$
\begin{equation} \label{gbalance3}
    X= \oint du \langle  J^{bc} (u) \ldots c(z_1) ...c(z_{N_c})b(y_1)...b(y_{N_b}) \rangle.
\end{equation}
Here, the ellipsis $\ldots$ stands for other types of fields. The contour integral encircles all the insertion points $z_i$ and $y_i$. Using the OPE
\be\label{gbalance4}
    J^{bc}(u) c(0) = -c(0)/u\,,\quad   J^{bc}(u) b(0) = b(0)/u\,,
\ee
we conclude that the insertion of such an operator effectively counts the number of $c $ and $b$ fields:
\begin{equation}\label{gbalance5}
    X= (N_c-N_b)   \langle  ...c(z_1) ...c(z_{N_c})b(y_1)...b(y_{N_b})\rangle.
\end{equation}
Now, deforming the contour to infinity and using the transformation law of the ghost current
\be   \label{gbalance6}
    J^{bc}(u)\to- J^{bc}(1/u)/u^2+3/u\,,\quad u \to 1/u,
\ee
we find that the same integral becomes
\begin{equation}  \label{gbalance7}
\begin{aligned}
    X= &  \oint_{\infty} du  [ \langle (-J(1/u)/u^2+3 /u   )   ...c(1/z_1) ...c(1/z_{N_c})b(1/y_1)...b(1/y_{N_b})\rangle]   \\&=3  \langle  ...c(1/z_1) ...c(1/z_{N_c})b(1/y_1)...b(1/y_{N_b})\rangle \\ & =
     3\langle  ...c(z_1) ...c(z_{N_c})b(y_1)...b(y_{N_b})\rangle.
    \end{aligned}
\end{equation}
which together with (\ref{gbalance5}) prove the first rule. For other rules, the proof follows the same rationale. 

Similarly, for the $\beta \gamma $ system, we have the ghost current $J^{\beta \gamma}=-:\beta \gamma:=-\partial\phi$ which transforms as 
\be
    J^{\beta \gamma}(u)\to- J^{\beta \gamma}(1/u)/u^2-2/u\,,\quad u \to 1/u\,.
\ee
Using the OPEs of  $J^{\beta \gamma}$ with $\beta, \gamma, \delta(\gamma)$, and applying the same logic 
 as we used for the $bc$ system above, one arrives at the second rule (\ref{gbalance2}).

 For the Ramond sector, we proceed analogously. In this case, we require the OPEs of the ghost current $J^{\beta\gamma}$ with the field $\sigma$ introduced above. It is convenient to work in the bosonized representation, where $\sigma = e^{-\phi/2}$. For an arbitrary exponential field of the form $e^{l\phi}$, its conformal dimension is given by $-l(l+2)/2$. In particular, for our purposes, we introduce the field $\sigma_2 = e^{\phi/2}$, which has conformal dimension $-5/8$. This allows one to construct fields (satisfying the constraint~\eqref{delta-tot}) from $\sigma_2$, $c$, and suitable descendants from the Liouville and matter sectors involving modes such as $G_{-1}$ and $L_{-1}$. For any exponential field, we have the OPE:

\begin{equation}
    J^{\beta \gamma} (u) e^{l \phi(0)} = \frac{l}{u} e^{l \phi(0)}.
\end{equation}
From this OPE, it follows that the modified ghost balance rule for correlation functions in the Ramond sector, which include insertions of $\sigma$ and $\sigma_2$, must take the form
\begin{equation}
\label{Nbg_gen}
     N_{\delta(\gamma)} - N_{\delta (\beta)}+ N_{\beta}- N_{\gamma} + N_{\sigma}/2- N_{\sigma_2}/2 =2.
\end{equation}
Restricting the rule (\ref{Nbg_gen}) to the case where only the fields $\delta(\gamma)$, $\sigma$, and $\sigma_2$ appear in the $\beta\gamma$ sector of the correlation function, we have
\begin{equation} \label{sigmarule}
         N_{\delta(\gamma)} + N_{\sigma}/2- N_{\sigma_2}/2 =2.
\end{equation}
A final remark is in order. In view of equation~\eqref{RamondPhys}, the presence of $\sigma_2$ appears to be essential. For instance, if $N_{\delta(\gamma)} = 2$, one must include pairs of $\sigma$ and $\sigma_2$ in higher-point functions. This requirement was discussed in \cite{Friedan:1985ge} in the context of vertex operators. For the three-point correlation function, the only possibility is given by (\ref{3ptnumber}). For the four-point correlation numbers we have a couple of options, their computation remains an interesting open question that we leave for future work.
\subsection{Ground ring operators}
Similarly to bosonic case \cite{Zamolodchikov:2003yb} and the NS sector \cite{Belavin:2008vc}, in addition to the physical states discussed above, the spectrum in the Ramond sector also contains ``discrete states'' generated by operators $\mathbb{O}_{m,n}$ 
\be \label{groundringop1}
    \mathbb{O}_{m,n}=\bar{H}_{m,n}H_{m,n} \mathbb{Y}^{R}_{m,n}\bar{\sigma}\sigma\,,\quad \mathbb{Y}^{R}_{m,n}=\left(\Theta_{m,n}^{-}R_{m,n}^{+}+i\Theta_{m,n}^{+}R_{m,n}^{-}\right)\,,
\ee
where the operators $H_{mn}$ are operators of dimension $\frac{mn}{2}-1$  built from super-Virasoro generators in Liouville and matter sectors, as well as ghost fields. 
They are fixed by requirement that $\mathbb{O}_{m,n}$ is BRST-closed, and therefore
\be  \label{groundringop2}
    \mathcal{Q}H_{m,n}\mathbb{Y}_{m,n}^{R}\sigma=
    \frac{1}{2}\left(1-\frac{1}{(\beta^{M}_{m,n})^2}G_0^{L}G_0^{M}\right)D_{m,n}^{L}
    \mathbb{Y}_{m,n}^{R}c\sigma.
\ee
For the simplest ground ring operator in the Ramond sector, namely $\mathbb{O}_{1,2}$, we find that the corresponding operator $H_{1,2}$ is given by
\be \label{groundringop3}
    H_{1,2}=\frac{1}{2}+\frac{b^2}{1-2b^2}G_0^{M}\beta_{-1}c_1-\frac{b^2}{1+2b^2}G_0^{L}\beta_{-1}c_1+\frac{4b^2}{(1-2b^2)(1+2b^2)}G_0^{L}G_0^{M}.
\ee
Taking advantage of the Higher Equations of Motion (\ref{HEM})
and the fact that $|\mathbb{U}_{m,n}\rangle$ is an eigenstate of the product $G_0^{L}G_0^{M}$
\be
\label{G0MLeigen}
G_0^{L}G_0^{M}\bar{D}^{L}_{m,n}D^{L}_{m,n}{(\mathbb{Y}^{R}_{m,n})}^\prime=B_{m,n}G_0^{L}G_0^{M}\mathbb{U}^{R}_{m,-n}=
iB_{m,n}\beta^{M}_{m,n}\beta^{L}_{m,-n}\mathbb{U}_{m,-n}^{R},
\ee
where
\be
    {(\mathbb{Y}^{R}_{m,n})}^\prime=\frac{\partial}{\partial a}\left(\Theta^{-}_{m,n}R^{+}_a
    +i\Theta^{+}_{m,n}R^{-}_a\right)\Big|_{a=a_{m,n}}\,,
\ee
we derive a convenient representation for the physical fields
\be
\label{RQQrep}
\bar{\mathcal{Q}}\mathcal{Q}\mathbb{O}_{m,n}^{\prime}=B_{m,n}\mathbb{R}_{m,-n}.
\ee
Acting on both sides of equation (\ref{RQQrep}) with $\bar{b}_{-1}b_{-1}$
we obtain the following corollary
\be \label{totalderivativeeq}
    B_{m,n}\mathbb{U}^{R}_{m,-n} \bar{\sigma}\sigma=
    \left(\bar{\partial}-\bar{\mathcal{Q}}\,\bar{b}_{-1}\right)
    (\partial -\mathcal{Q}b_{-1})\mathbb{O}^{\prime}_{m,n}
    =\bar{\partial}\partial \mathbb{O}^{\prime}_{m,n}+
    \text{BRST exact terms}
\ee
which can be used to facilitate evaluation of moduli integrals in four-point correlation numbers.

\subsection{Three-point correlation number from the ground ring operator $\mathbb{O}_{1,2}$}
In this section, we rederive the expression (\ref{3ptnumber}) by using the representation (\ref{RQQrep}) for one of the Ramond fields in the correlation number. This derivation serves as a consistency check of (\ref{3ptnumber}) itself and of the validity of  (\ref{RQQrep}).  The idea follows the same logic explained in appendix \ref{apNS3ptcn} (where we obtain, from ground ring operators, the three-point correlation number of three NS fields). Thus, for the first field in  (\ref{3ptnumber}) we have
\begin{equation} \label{Q3ptnum1}
    \mathbb{R}_{1,-2}= \frac{1}{ B_{1,2}}\bar{\mathcal{Q} }\mathcal{Q}\mathbb{O}'_{1,2}.
\end{equation}
We aim to compute 
\begin{equation} \label{Q3ptnum2}
 I_3^{R}  (3b/2, a,a-b/2) = \langle  \mathbb{R}_{1,-2} (z_1)  \mathbb{R}_a(z_2)  \mathbb{W}_{a-b/2}(z_3)   \rangle=  \frac{1}{B_{1,2}} \langle \bar{\mathcal{Q} }\mathcal{Q}\mathbb{O}'_{1,2}   \mathbb{R}_a(z_2)  \mathbb{W}_{a-b/2}(z_3) \rangle.
\end{equation}
By computing $\mathbb{O}_{1,2}$ explicitly from (\ref{groundringop1}) and (\ref{groundringop3}),
and considering the OPE structure of $ \langle  V_a   R^{\epsilon_1}_{a_1} R^{\epsilon_2}_{a_2} \rangle$ determined by  (\ref{apOPERRV}), along with the special OPEs in Liouville and matter sectors 
\begin{equation} \label{Q3ptnum6}
\begin{aligned}
    R_{-b/2}^{\epsilon} (z)R^{\epsilon}_{a_1}(0) =  & z^{\Delta^L(a_1+b/2)- \Delta^L(-b/2)-\Delta^L(a_1)} \mathbf{C}^{L, \epsilon}_{ [-b/2] [a_1] (a_1+b/2)} V_{a_1+b/2}(0)+\\& z^{\Delta^L(a_1-b/2)- \Delta^L(-b/2)-\Delta^L(a_1)} \mathbf{C}^{L,{\epsilon}}_{ [-b/2] [a_1] (a_1-b/2)} V_{a_1-b/2}(0),
\end{aligned}
\end{equation}
\begin{equation} \label{Q3ptnum7}
\begin{aligned}
    \Theta_{b/2}^{\epsilon} (z)\Theta^{\epsilon}_{a_1}(0) =  & z^{\Delta^M(a_1+b/2)- \Delta^M(b/2)-\Delta^M(a_1)} \mathbf{C}^{M,\epsilon}_{ [b/2] [a_1] (a_1+b/2)} \Phi_{a_1+b/2}(0)+\\& z^{\Delta^M(a_1-b/2)- \Delta^M(b/2)-\Delta^M(a_1)} \mathbf{C}^{M,{\epsilon}}_{ [b/2] [a_1] (a_1-b/2)} \Phi_{a_1-b/2}(0),
    \end{aligned}
\end{equation}
one finds that in the OPE  $\mathbb{O}_{1,2} (z) \mathbb{R}_a(0)$ the terms that may  yield non-zero contributions in (\ref{Q3ptnum2}) are 

\begin{equation}  \label{Q3ptnum7}
    \mathbb{O}_{1,2}(z) \mathbb{R}_a(0) =   A_{a+b/2} \mathbb{W}_{a+b/2}(0)+  A_{a-b/2}  \mathbb{W}_{a-b/2}(0)+...,
\end{equation}
where we have taken into account that
\begin{equation}
\sigma(z)\sigma(0)= z^{-1/4} \delta(\gamma(0)),
\end{equation}
and the coefficients $A$ are given by
\begin{equation} \label {coefAa}
    A_{a+b/2}=\frac{1}{2} \mathbf{C}^{M, -} _ {  [b/2] [a-b] (a-b+b/2)   } \mathbf{C}^{L, +} _ {  [-b/2] [a] (a+b/2)   }- \frac{1}{2}\mathbf{C}^{M,+} _ {  [b/2] [a-b] (a-b+b/2)   } \mathbf{C}^{L,-} _ {  [-b/2] [a] (a+b/2)   },
\end{equation}
\begin{equation} \label{coefAaM}
        A_{a-b/2}= \frac{1}{2}\mathbf{C}^{M, -} _ {  [b/2] [a-b] (a-b-b/2)   } \mathbf{C}^{L, +} _ {  [-b/2] [a] (a-b/2)   }- \frac{1}{2}\mathbf{C}^{M,+} _ {  [b/2] [a-b] (a-b-b/2)   } \mathbf{C}^{L,-}_ {  [-b/2] [a] (a-b/2)   }.
\end{equation}
To  compute  (\ref{coefAa}, \ref{coefAaM}) one needs to obtain the special structure constant in Liouville sector. These can be computed using the screening calculus \cite{Dotsenko:1984nm} or directly from (\ref{Ramondstructconst}) by evaluating the appropriate limit (see section 2 and appendix B of \cite{Belavin:2007gz} for the description of both approaches). The result from (\ref{Ramondstructconst}) reads

\begin{equation} \label{specialLstrucc}
\begin{aligned} 
& \mathbf{C}^{L,\epsilon} _ {  [-b/2] [a] (a+b/2)   }  =    \frac{2 \pi  b^2 u \gamma \left(\frac{1}{2} \left(b^2+1\right)\right) \gamma \left(b \left(a-\frac{b}{2}\right)\right)}{\gamma \left(a b+\frac{1}{2}\right)},\\&
\frac{\mathbf{C}^{L,\epsilon} _ {  [-b/2] [a] (a-b/2)   }}{2}= \epsilon .
 \end{aligned}
 \end{equation}
 Now, following the appendix \ref{apNS3ptcn}, to compute the OPE  $\mathbb{O'}_{1,2}(z) \mathbb{R}_a(0) $ we multiply (\ref{Q3ptnum7}) by the scalar field $\varphi(z_1)$. With this at hand, we write the rightmost part of (\ref{Q3ptnum2}) 
 \begin{equation}
      I_3^{R}  (3b/2, a,a-b/2) = \frac{1}{ B_{1,2}} \frac{2}{ 2 \pi i }\oint_{z_1} du \frac{1}{2 \pi i}\oint_{\bar{z}_1} d\bar{u}  \langle \bar{j}_{\mathcal{Q}}(\bar{u})j_{\mathcal{Q}}(u) \mathbb{O}'_{1,2}(z_1)\mathbb{R}_{a}(z_2)\mathbb{W}_{a-b/2}(z_3) \rangle.
 \end{equation}
Considering (\ref{deri3pt7}) and computing the three-point function, similar to (\ref{deri3pt9}), which arises from the Liouville sector, we obtain that
  \begin{equation}
  \begin{aligned}
     I_3^R(3b/2,a,a-b/2) &=        \frac{2}{   B_{1,2}}   A_{a-b/2} G_{NS}(a-b/2) (Q/2- (a-b/2))^2 \\& =\frac{ 2 \mathbf{C}^{M,+} _ {  [b/2] [a-b] (a-b-b/2)   }  G_{NS} (a-b/2)}{  B_{1,2}} (Q/2- (a-b/2))^2,
    \end{aligned}
 \end{equation}
 which agrees with equation (\ref{3ptnumber}) for $(a_1,a_2,a_3)= (3b/2,a,a-b/2)$. Similarly, for the case $a_3=a+b/2$ one obtains the same type of relation.

\section{Conclusion}
In this work, we analyzed the construction of physical fields in the Ramond sector of super minimal Liouville gravity. We derived the the three-point correlation number involving two physical fields in the Ramond sector (\ref{RamondPhys}) and one physical field in the NS sector (\ref{W}), the result is given by relations (\ref{3ptnumber}, \ref{NRlegfactor}, \ref{ROmegafactor}), which constitute part of the main results of the paper. Another part of our results concerns the derivation of the ground ring operators (\ref{groundringop1}, \ref{groundringop3}) and the more general expressions (\ref{RQQrep}, \ref{totalderivativeeq}), which play a fundamental role in the analytical treatment of SMLG. The result (\ref{3ptnumber}) was subsequently verified through a direct computation using the expression for the ground ring operator. In section~\ref{ghostnumberrules}, we also analyzed the constraints on the types of fields that can appear in higher-point correlation numbers. The findings of this paper provide, in principle, a method for performing analytic computations of the moduli integrals that arise in the construction of multi-point correlation functions involving Ramond field insertions. This is one of our next objectives.

Another direction is related to the investigation of the dual approach, which has not yet been explored on the level of correlation numbers.  However, some arguments favor the hypothesis that the results are essentially the same as those of the bosonic version, see~\cite{Seiberg:2003nm}. This hypothesis can be easily tested in the NS sector, and we plan to extend the investigation to the Ramond sector as well. To do so, one needs to explicitly compute certain correlators involving Ramond fields.

\appendix

\input{shifts}
\section{NS three-point correlation number} \label{apNS3ptcn}
In this section, we rederive the three-point correlation number in the NS sector by using the contour integral of the BRST charge. The three-point correlation number of NS physical fields is given by (here $z_1 = 0$, $z_2 = 1$, $z_3 = \infty$; since these are physical fields, the dependence on coordinates drops out):
\begin{equation} \label{deri3pt1}
\begin{aligned}
I_3(a_1,a_2,a_3) & =  \langle   \tilde{\mathbb{W}}_{m,-n}(z_{1}) \mathbb{W}_{a_{2}}(z_{2}) \mathbb{W}_{a_{3}}(z_{3}) \rangle \\& =\frac{1}{B_{m,n}} \frac{1}{ (2 \pi i)^2} \oint _{z_1} d^2z      \langle \bar{j}_{\mathcal{Q}}(\bar{z}) j_{\mathcal{Q}}(z)   \mathbb{O}'_{m,n}(z_1)\mathbb{W}_{a_{2}}(z_{2}) \mathbb{W}_{a_{3}}(z_{3}) \rangle      .
\end{aligned}
\end{equation}
We can rewrite $\tilde{\mathbb{W}}_{m,-n}$ as $\tilde{\mathbb{W}}_{a_1}$. Let us analyze a simple case, when $m = 1$, $n = 3$ (hence $a_1 = 2b$), and $a_2 = a_3 = a$. The expression $I_3(2b, a, a)$ is determined by equation~(\ref{purNSI3}). Let us recheck this by computing the rightmost side of the expression above. Thus, we want to compute

\begin{equation} \label{deri3pt2}
  I_3=  \frac{1}{B_{1,3}} \frac{1}{2 \pi i}\oint _{z_1} dz\frac{1}{2 \pi i}\oint_{\bar{z}_1} d\bar{z}       \langle  \bar{j}_{\mathcal{Q}}(\bar{z}) j_{\mathcal{Q}}(z) \mathbb{O}'_{1,3}(z_1)\mathbb{W}_{a}(z_{2}) \mathbb{W}_{a}(z_{3}) \rangle .     
\end{equation}
Recall (see  \cite{Belavin:2008vc} for details) that  
\begin{equation}  \label{deri3pt3}
\begin{aligned}
\mathbb{O}_{13}(x) =\, & \Phi_{13}^{\prime}(x)\, V_{13}(x)
- \Phi_{13}(x)\, V_{13}^{\prime}(x) +\\
& \overbrace{-
\Psi_{13}(x)\, \Lambda_{13}(x) +}^{\text{term contributing to } \mathbb{W}_a \text{ in the OPE $ \mathbb{O}_{1,3}\mathbb{W}_a$}}\\&
+ \left[b^2 : \beta(x)\gamma(x): + 2 b^2 : b(x) c(x): \right] \Phi_{13}(x)\, V_{13}(x)
 \\
& - b^2 \beta(x)\, c(x)\, \Psi_{13}(x)\, V_{13}(x)
- b^2 \beta(x)\, c(x)\, \Phi_{13}(x)\, \Lambda_{13}(x),
\end{aligned}
\end{equation}
where $\Lambda_{13}=G^{L}_{-1/2}V_{13}, \Psi_{13}=G^{M}_{-1/2}\Phi_{13}$ . In the OPE $\mathbb{O}_{1,3}\mathbb{W}_a$ the term proportional to $\mathbb{W}_a$ arises from the third term of the expansion above. Thus we have
\begin{equation} \label{deri3pt4}
    \mathbb{O}_{1,3}(z_1) \mathbb{W}_{a}(z_2)= \tilde{\mathbf{C}}^M_0(a-b)\tilde{\mathbf{C}}^L_{0}(a) \mathbb{W}_a(z_2)+... \quad,
\end{equation}
where $ \tilde{\mathbf{C}}^M_0(a-b), \tilde{\mathbf{C}}^L_{0}(a) $ are special matter and Liouville structure constants given by 
\begin{equation}
\tilde{\mathbf{C}}_0^{\mathrm{L}}(a)=\frac{2 \pi i \mu \gamma\left(a b-b^2\right)}{\gamma\left(-b^2\right) \gamma(a b)}, 
\end{equation}
\begin{equation}
\tilde{\mathbf{C}}_0^{\mathrm{M}}(a)=i b^{-2} \gamma\left(\frac{b Q}{2}\right)\left(\frac{\gamma\left(1-b^2\right) \gamma\left(b^2 / 2-1 / 2\right)}{\gamma\left(b^2-1\right) \gamma\left(3 b^2 / 2-1 / 2\right)}\right)^{1 / 2} \frac{\gamma\left(a b+b^2\right)}{\gamma(a b)}.
\end{equation}
Since $V_a(z)=e^{a \varphi(z)}$, hence $V'_a=\partial_a V_a(z)= \varphi(z) V_a(z)$. Thus
\begin{equation} \label{deri3pt5}
      \mathbb{O'}_{1,3}(z_1) \mathbb{W}_{a}(z_2) = \varphi (z_1) \left(\tilde{\mathbf{C}}^M_0(a-b)\tilde{\mathbf{C}}^L_{0}(a)\mathbb{ W}_a(z_2) \right).
\end{equation}
Inserting this OPE into (\ref{deri3pt2}) gives
\begin{equation} \label{deri3pt6}
      I_3=  \frac{ \tilde{\mathbf{C}}^M_0(a-b)\tilde{\mathbf{C}}^L_{0}(a)}{B_{1,3}} \frac{1}{2 \pi i}\oint _{z_1} dz\frac{1}{2 \pi i}\oint_{\bar{z}_1} d\bar{z}       \langle  \bar{j}_{\mathcal{Q}}(\bar{z}) j_{\mathcal{Q}}(z) \varphi(z_1)\mathbb{W}_{a}(z_{2}) \mathbb{W}_{a}(z_{3}) \rangle .   
\end{equation}
The action of $j_{\mathcal{Q}}$ on $\mathbb{W}_a$ gives no contribution since $\mathbb{W}_a$ is physical. The non-zero contribution comes from the OPE $j_{\mathcal{Q}}(z)\, \varphi(z_1)$. Thus, we have
\begin{equation} \label{deri3pt7}
    j_{\mathcal{Q}} (z) \varphi (z_1)= \left( c(z) T(z)+...\right)\varphi(z_1)= c(z)\frac{\varphi' (z)}{z-z_1}. 
\end{equation}
Therefore
\begin{equation}  \label{deri3pt8}
      I_3=  \frac{ \tilde{\mathbf{C}}^M_0(a-b)\tilde{\mathbf{C}}^L_{0}(a)}{B_{1,3}} \frac{1}{2 \pi i}\oint _{z_1} dz\frac{1}{2 \pi i}\oint_{\bar{z}_1} d\bar{z}    \frac{1}{|z-z_1|^2}   \langle \bar{c}(\bar{z}) c(z)  \partial _z\partial _{\bar{z}}\varphi(z)\mathbb{W}_{a}(z_{2}) \mathbb{W}_{a}(z_{3}) \rangle .   
\end{equation}
Evaluating the integral gives
\begin{equation} \label{deri3pt9}
      I_3=  \frac{ \tilde{\mathbf{C}}^M_0(a-b)\tilde{\mathbf{C}}^L_{0}(a)}{B_{1,3}}   \langle  \bar{c}(\bar{z}_1) c(z_1) \varphi'(z_1)\mathbb{W}_{a}(z_{2}) \mathbb{W}_{a}(z_{3}) \rangle .   
\end{equation}
Recall that $\varphi'$ is a primary field with conformal dimensions $(1,1)$. Evaluating the corresponding three-point function in the Liouville sector, we arrive at the final result:

\begin{equation} \label{deri3pt10}
    I_3= \frac{ 2\tilde{\mathbf{C}}^M_0(a-b)\tilde{\mathbf{C}}^L_{0}(a)}{B_{1,3}}  G_{NS}(a)   (- a+Q/2)^2,
\end{equation}
where
$G_{NS}$ arises from $ \langle  \mathbb{W}_{a}  \mathbb{W}_{a}\rangle $ which is determined by the reflection coefficient (\ref{refcoefRN}). The above result agrees with (\ref{purNSI3}).


\clearpage
\bibliographystyle{JHEP} 
\bibliography{bib} 
\end{document}

%% file: shifts.tex
\section{Shift relations in R sector} \label{apShiftRelation}
The purpose of this section is to derive the  three-point constants involving two fields in the Ramond sector. We use the approach first outlined in \cite{Zamolodchikov:2005}, namely deriving the shift relations for the whole supersymmetric MLG, which, as will be demonstrated below, are considerably simpler  than similar relations in either Liouville or matter sector considered separately.

It is convenient to introduce a special notation for the following linear combination of the structure constants:
\be \label{apeshiftrela1}
    \mathbf{X}^{\sigma}(\alpha_1,\alpha_2|\alpha_3)=\mathbf{C}^{+}_{[\alpha_1][\alpha_2](\alpha_3)}+\sigma \mathbf{C}^{-}_{[\alpha_1][\alpha_2](\alpha_3)}\,.
\ee
The crossing symmetry of a four-point function involving a degenerate field $R_{1,2}$ imposes a system of relations on the structure constants \cite{Poghossian:1996agj}.
Introducing the following auxiliary functions
\be
    \rho_{\epsilon}=-\frac{\Gamma^2(c_{\epsilon})\gamma(1-a_{\epsilon})\gamma(1-b_{\epsilon})}
{\Gamma^2(2-c_{\epsilon})\gamma(c_{\epsilon}-a_{\epsilon})\gamma(c_{\epsilon}-b_{\epsilon})}\,,
\ee
\be \label{apeshiftrela2}
    h_{\epsilon}=\frac{\gamma(c_{\epsilon})\gamma(c_{\epsilon}-a_{\epsilon}-b_{\epsilon})}
    {\gamma(c_{\epsilon}-a_{\epsilon})\gamma(c_{\epsilon}-b_{\epsilon})}\,,\quad
    \tilde{h}_{\epsilon}=\frac{\gamma(c_{\epsilon})\gamma(a_{\epsilon}+b_{\epsilon}-c_{\epsilon})}
    {\gamma(a_{\epsilon})\gamma(b_{\epsilon})}\,,
\ee
where $a_{\epsilon}$, $b_{\epsilon}$ and $c_{\epsilon}$ are functions of Liouville momenta $\alpha_i$
\begin{eqnarray}
    a_{\epsilon}&=&\frac{1}{4}\left(1+b\epsilon(3Q-2\alpha_1-2\alpha_2-2\alpha_3)\right),\\
    b_{\epsilon}&=&\frac{1}{4}\left(1+b\epsilon(Q-2\alpha_1-2\alpha_2+2\alpha_3)\right),\\
    c_\epsilon&=&\frac{1}{2}(1+b\epsilon(Q-2\alpha_2)),
\end{eqnarray}
     we can summarize relevant relations as follows:
\begin{eqnarray}
    h_{+}(\alpha_1,\alpha_2,\alpha_3)g_{+}(\alpha_1,\alpha_2,\alpha_3) &=& \mathbf{C}_{-}^{+}(\alpha_2+\frac{b}{2})\mathbf{X}^{-}(\alpha_1,\alpha_2+\frac{b}{2}|\alpha_3),\\
    \tilde{h}_{-}(\alpha_1,\alpha_2,\alpha_3)g_{-}(\alpha_1,\alpha_2,\alpha_3)&=&\mathbf{C}_{-}^{+}(\alpha_2+\frac{b}{2})\mathbf{X}^{+}(\alpha_1,\alpha_2+\frac{b}{2}|\alpha_3),\\
    \tilde{h}_{+}(\alpha_1,\alpha_2,\alpha_3)g_{+}(\alpha_1,\alpha_2,\alpha_3)&=&\mathbf{C}_{+}^{+}(\alpha_2-\frac{b}{2})\mathbf{X}^{-}(\alpha_1,\alpha_2-\frac{b}{2}|\alpha_3),\\
    h_{-}(\alpha_1,\alpha_2,\alpha_3)g_{-}(\alpha_1,\alpha_2,\alpha_3)&=&\mathbf{C}_{+}^{+}(\alpha_2-\frac{b}{2})\mathbf{X}^{+}(\alpha_1,\alpha_2-\frac{b}{2}|\alpha_3),\\
    \frac{g_{+}(\alpha_1+\frac{b}{2},\alpha_2,\alpha_3)}
    {g_{-}(\alpha_1-\frac{b}{2},\alpha_2,\alpha_3)}&=&
    \frac{\rho_{-}(\alpha_1-\frac{b}{2},\alpha_2,\alpha_3)(\beta_{\alpha_1+\frac{b}{2}}+\beta_{-\frac{b}{2}})^2\mathbf{C}^{+}_{-}(\alpha_1+\frac{b}{2})}
    {\rho_{+}(\alpha_1+\frac{b}{2},\alpha_2,\alpha_3)(\beta_{\alpha_1-\frac{b}{2}}-\beta_{-\frac{b}{2}})^2\mathbf{C}^{+}_{+}(\alpha_1-\frac{b}{2})},\\
    \frac{g_{+}(\alpha_1-\frac{b}{2},\alpha_2,\alpha_3)}
    {g_{-}(\alpha_1+\frac{b}{2},\alpha_2,\alpha_3)}&=&
    \frac{\mathbf{C}_{+}^{+}(\alpha_1-\frac{b}{2})}{\mathbf{C}_{-}^{+}(\alpha_1+\frac{b}{2})},
\end{eqnarray}
where, in contrast with section \ref{superMLG}, we have used GMM-like normalization 
\be \label{apeshiftrela3}
    \langle R_{\alpha_1}^{\epsilon_1}(x)R_{\alpha_2}^{\epsilon_2}(0)\rangle=\delta(\alpha_1-\alpha_2)\delta_{\epsilon_1,\epsilon_2}(x\bar{x})^{-2\Delta(\alpha_1)},
\ee
and a shorthand notation for special structure constants (\ref{specialstrc})
\be
    \mathbf{C}^{\epsilon}_{\sigma}(\alpha)=
    \mathbf{C}^{\epsilon}_{[\alpha][-\frac{b}{2}](\alpha+\frac{\sigma b}{2})}\,.
\ee
Shifting the first and the second arguments as necessary we can exclude the unknown functions $g_{\pm}$ and obtain the following functional equations for the constants $\mathbf{X}^{\pm}$
\be \label{apeshiftrela4}
\begin{split}
    &\frac{\mathbf{X}^{+}(\alpha_1+b,\alpha_2|\alpha_3)}{\mathbf{X}^{+}(\alpha_1-b,\alpha_2|\alpha_3)}=
\frac{\gamma(\alpha_1 b +\frac{b^2}{2}+1)\gamma(\alpha_1 b-\frac{b^2}{2}+1)}
    {\gamma(-\alpha_1 b+\frac{3}{2}b^2+2)\gamma(2-\alpha_1 b+\frac{b^2}{2})}\\
    &\times\frac{\gamma(-\frac{1}{2}\alpha_{1-2-3}b-\frac{b^2}{2})\gamma(\frac{1}{2}-\frac{1}{2}\alpha_{1-2-3}b)
\gamma(\frac{1}{2}-\frac{1}{2}\alpha_{1-2+3})\gamma(\frac{1}{2}-\frac{1}{2}\alpha_{1+2-3}b+\frac{1}{2}b^2)
    }
    {\gamma(\frac{1}{2}\alpha_{1+2+3}b-\frac{b^2}{2}-\frac{1}{2})\gamma(\frac{1}{2}\alpha_{1+2+3}b-b^2)
    \gamma(\frac{1}{2}\alpha_{1+2-3}b)\gamma(\frac{1}{2}\alpha_{1-2+3}b-\frac{1}{2}b^2)
    }\\
    &\times
    \frac{(\alpha_1 b-\frac{1}{2})(\alpha_1 b-b^2-\frac{1}{2})\gamma(\alpha_1 b-\frac{1}{2})\gamma(\alpha_1 b-b^2-\frac{1}{2})}
    {\gamma(\alpha_1 b -\frac{b^2}{2})\sqrt{\gamma(\alpha_1 b+\frac{b^2}{2})\gamma(\alpha_1 b-\frac{3b^2}{2})}}
    \frac{{\left({\alpha_1} b - \frac{b^{2}}{2} - 1\right)}^{2} {\left({\alpha_1} b - \frac{3}{2} b^{2} - 1\right)}^{2}}
    {{\left( {\alpha_1}b + \frac{1}{2}b^2\right)}^{2} {\left( {\alpha_1}b - \frac{1}{2}b^2\right)}^{2}},
\end{split}
\ee 
\be  
\label{xmratDeltaN}
\begin{split}
    &\frac{\mathbf{X}^{-}(\alpha_1+b,\alpha_2|\alpha_3)}{\mathbf{X}^{-}(\alpha_1-b,\alpha_2|\alpha_3)}=
    \frac{\gamma(\alpha_1 b +\frac{b^2}{2}+1)\gamma(\alpha_1 b-\frac{b^2}{2}+1)}
    {\gamma(-\alpha_1 b+\frac{3}{2}b^2+2)\gamma(2-\alpha_1 b+\frac{b^2}{2})}\\
    &\times\frac{\gamma(\frac{1}{2}-\frac{1}{2}\alpha_{1-2-3}b-\frac{1}{2}b^2)\gamma(\frac{1}{2}-\frac{1}{2}\alpha_{1-2+3}b+\frac{1}{2}b^2)
    \gamma(\frac{1}{2}-\frac{1}{2}\alpha_{1+2-3}b)\gamma(-\frac{1}{2}\alpha_{1-2-3}b)}
    {\gamma(\frac{1}{2}\alpha_{1+2+3}b-\frac{1}{2}b^2)\gamma(\frac{1}{2}\alpha_{1+2+3}-b^2-\frac{1}{2})
    \gamma(\frac{1}{2}\alpha_{1+2-3}b-\frac{1}{2}b^2)\gamma(\frac{1}{2}\alpha_{1-2+3}b)}\\
    &\times
    \frac{(\alpha_1 b-\frac{1}{2})(\alpha_1 b-b^2-\frac{1}{2})\gamma(\alpha_1 b-\frac{1}{2})\gamma(\alpha_1 b-b^2-\frac{1}{2})}
    {\gamma(\alpha_1 b -\frac{b^2}{2})\sqrt{\gamma(\alpha_1 b+\frac{b^2}{2})\gamma(\alpha_1 b-\frac{3b^2}{2})}}
    \frac{{\left({\alpha_1} b - \frac{b^{2}}{2} - 1\right)}^{2} {\left({\alpha_1} b - \frac{3}{2} b^{2} - 1\right)}^{2}}
    {{\left( {\alpha_1}b + \frac{1}{2}b^2\right)}^{2} {\left( {\alpha_1}b - \frac{1}{2}b^2\right)}^{2}}.
\end{split}
\ee
Note that the transformation $(\alpha,b)\mapsto (i(\alpha-b),-ib)$ maps the ratio of the structure constants of the supersymmetric Liouville theory into a ratio of the structure constants of a GMM as follows
\be \label{apeshiftrela6}
    \frac{\mathbf{X}^{\sigma}(\alpha_1+b,\alpha_2|\alpha_3)}
    {\mathbf{X}^{\sigma}(\alpha_1-b,\alpha_2|\alpha_3)}
    \mapsto
    \frac{\mathbf{X}_M^{\sigma}(\alpha_1-2b,\alpha_2|\alpha_3)}
    {\mathbf{X}^{\sigma}_M(\alpha_1,\alpha_2|\alpha_3)}\,.
\ee
Then, restoring the normalization of section \ref{superMLG} in the Liouville sector
and performing the substitution in (\ref{apeshiftrela4}, \ref{xmratDeltaN}) one finds that for the ratios of the following  products of the structure constants 
\be
\label{XMLGdef}
    \mathbf{X}^{\pm}_{MLG}(\alpha_1,\alpha_2|\alpha_3)=
    \mathbf{X}^{\pm}_{L}(\alpha_1,\alpha_2|\alpha_3)
    \mathbf{X}^{\mp}_{M}(\alpha_1-b,\alpha_2-b|\alpha_3-b)
\ee
the dependence on the parameters $\alpha_2,\alpha_3$ cancels out:
\be \label{apRshiftRela}
    \frac{\mathbf{X}^{\pm}_{MLG}(\alpha_1+b,\alpha_2|\alpha_3)}
    {\mathbf{X}^{\pm}_{MLG}(\alpha_1-b,\alpha_2,\alpha_3)}=\frac{N_R(\alpha_1+b)}{N_R(\alpha_1-b)}\,,
\ee
where by $N_R$ we have denoted the Ramond sector leg factor:
\be
 N_R(a)=\sqrt{\gamma \left(\frac{a}{b}-\frac{1}{2 b^2}\right) \gamma \left(a b-\frac{b^2}{2}\right)} \left(\pi  \mu \gamma \left(\frac{1}{2} \left(b^2+1\right)\right)\right)^{-\frac{a}{b}}\,,
\ee
The symmetry of the three-point function w.r.t. permutation of the two Ramond fields implies that MLG structure constants $\mathbf{X}^{\pm}_{MLG}$  must admit the following factorized form
\be
    \mathbf{X}^{\pm}_{MLG}(a_1,a_2|a_3)=\Omega_{R}^{(\pm)}(b)N_{R}(a_1)N_R(a_2)N_{NS}(a_3)\,.
\ee